\documentclass{elsart5p}

\usepackage{graphics}
 \usepackage{graphicx}

\usepackage{amssymb}
\newcommand{\beq}{\begin{equation}}
\newcommand{\eeq}{\end{equation}}
\newcommand{\beqa}{\begin{eqnarray}}
\newcommand{\eeqa}{\end{eqnarray}}

\newcommand{\qvec}{{\bf q}}
\newcommand{\Qvec}{{\bf Q}}
\newcommand{\OO}{{\overline \Omega}}

\begin{document}

\begin{frontmatter}



\title{Spectral signatures of critical charge and spin fluctuations in cuprates}


\author[label1]{M. Grilli},
\author[label1]{S. Caprara},
\author[label1]{C. Di Castro},
\author[label2]{T. Enss},
\author[label3]{R. Hackl},
\author[label3]{B. Muschler},
\author[label3]{and W. Prestel}

\address[label1]{Dipartimento di Fisica, Universit\`a di Roma ``Sapienza' e SMC-IMFM-CNR,
P.le A. Moro, 00185 Roma, Italy}
\address[label2]{Technische Universit\"at M\"unchen,
Physik Department T34,
James-Franck-Str., 
D-85747 Garching, Germany}
\address[label3]{Walther-Meissner-Institut,
Bayerische Akademie der Wissenschaften,
Walther-Meissner-Strasse 8,
86748 Garching,
Germany}

\begin{abstract}
We discuss how  Raman spectra of high temperature superconducting cuprates
are affected by nearly-critical spin and charge
collective modes, which are coupled to charge carriers near a stripe quantum critical point. 
We find that specific fingerprints of nearly-critical collective modes can be observed
and that the selectivity of Raman spectroscopy in momentum space may be exploited to distinguish
the spin and charge contribution. We apply our results to discuss the spectra of high-T$_c$c
superconducting cuprates finding that the collective modes should have masses with substantial
temperature dependence in agreement with their nearly critical character. Moreover spin modes 
have larger masses and are
more diffusive than charge modes indicating that in stripes the charge is nearly 
ordered, while spin modes
are strongly overdamped and fluctuating with high frequency.
\end{abstract}

\end{frontmatter}

\section{Introduction: The anomalous phase vs quantum criticality debate}
The origin of the anomalous metallic behavior of the superconducting cuprates
is still debated. One possibility is that the strongly correlated nature of these
doped Mott insulators, together with their nearly two-dimensional electronic structure,
gives rise to a stable metallic  phase with non-Fermi-liquid (non-FL) properties. Since the
early days of high-~T$_c$ superconductivity the proposal of a resonating valence bond (RVB)
state \cite{anderson1} was put forward with properties similar to those of the 
Luttinger liquid in one-dimensional systems \cite{anderson2}. Another proposal for a 
non-FL phase is related to the formation of nearly one-dimensional 
striped textures in the underdoped cuprates 
\cite{emerykivelson}. In this case the non-FL properties are structurally inherent to the 
 strongly inhomogenous electronic liquid state. In both cases, the non-FL character of
the cuprates is an intrinsic property of the ground state ``{\it per se}''.

A completely different point of view relates the anomalous properties to the proximity to
some second-order phase transition. In this case abundant long-range fluctuations 
of the order parameter can couple to the electrons strongly affecting their properties.
In particular, the instability line may extend to zero temperature by increasing
doping ending into a quantum critical point (QCP) \cite{varma,CDG,chubukov}. 
Thus the fluctuations acquire 
the dynamical character
of collective modes (CMs), which couple to the electrons mediating a {\it retarded} effective 
interaction. Owing to their nearly critical character, these CMs have low typical energy 
scales and retardation plays a relevant role. Anomalous properties then naturally arise
from the strong dynamical fluctuations. The two above alternative points of
view (namely the non-FL phase and the QCP scenario) precisely find a counterpart in the 
different nature of effective interactions: While in the non-FL phase the interactions
may well be instantaneous (like, e.g., a magnetic coupling J forming the singlet spin
liquid in the RVB phase), the QCP scenario relies on the presence of low-energy modes
mediating retarded interactions. These interactions also mediate pairing giving rise to
the so-called ``glue'' issue: understanding the retarded or non-retarded character of pairing
would shed light on the whole physics
of the cuprates. This explains the growing interest in the ``glue'' issue both from the
theoretical \cite{anderson3,scalapino} and experimental \cite{basov,vandermarel} side.

\section{The CO-QCP scenario}
The presence of non-FL behavior close to QCP's
in different correlated systems like the heavy fermions indirectly supports to
the relevance of this scenario for the cuprates. Yet the QCP line of reasoning 
allows for a variety of possible realizations. One possibility is that
the instability is (nearly) local in space and the instability gives rise to a (nearly) local QCP,
with the destruction of the FL state being induced by the formation of local singlets \cite{CFCT}.
In this case the critical fluctuations are expected to be rather structureless in momentum space.
On the other hand, strongly peaked fluctuations at zero momentum are expected if the system 
becomes unstable because of the formation of a spatially uniform state like in the case of a 
Pomeranchuk instability \cite{metzner}
or of a circulating current state breaking time reversal symmetry\cite{varma}.
The formation of a state breaking translational invariance naturally leads to fluctuations
strongly peaked at momenta related to the wavelength of the ordered state. This is the case of
the antiferromagnetic (AF) state, with a transition accompanied by critical spin fluctuations
strongly peaked at $\Qvec_s\approx (\pm \pi, \pm \pi)$. The AF state occurs in cuprates at 
very low doping and rapidly dies upon hole doping. Thus, disregarding disorder effects, possibly
leading to glassy phases at low temperature, one expects the QCP to be located at low
doping and the related physical effects (pseudogap, superconductivity, magnetic and transport
properties, and so on) to emanate at moderate-to-optimal doping starting
from this low-doping QCP \cite{chubukov}. 

Here we discuss evidences indicating that the physics of cuprates
is affected by the proximity to a charge-ordering (CO) QCP located 
near optimal doping \cite{CDG2}. 
The CO-QCP merges two relevant issues in the cuprates: the issue of charge
inhomogeneities and of quantum criticality. The first one arose theoretically
from the finding
of striped phases in mean-field treatments of the Hubbard model \cite{ZAANEN}.
Striped phases (as well as other inhomogeneous phases) can also be found when the
tendency to phase separation, common for strongly correlated systems is frustrated
by the long-range Coulomb force preventing segregation of charged carriers 
\cite{emerykivelson,RCDGBK,low}. Following this latter approach to stripes, the
CO instability was first analyzed in Ref. \cite{CDG} in the framework
of quantum criticality.  While in real systems true long-range charge order
is prevented by the competing formation of pairs, by disorder, and by the two-dimensional
structure of the copper-oxide planes, the physical properties of cuprates are strongly
affected by low-energy charge fluctuations on large domains. The putative instability
line would end in a QCP around optimal doping. In this framework 
the overdoped region corresponds to the
quantum disordered side of the phase diagram, the optimally doped region corresponds 
to the quantum critical and the underdoped region corresponds to the nearly ordered 
region. It is worth remarking that this ``missed'' QCP is not alternative to the
presence of strong spin fluctuations due to the AF state at low doping. On the contrary,
the intrinsically inhomogeneous CO state creates regions with lower hole density, where
spin degrees of freedom are important. Therefore charge fluctuations due 
naturally ``enslave'' spin fluctuations and bring the relevance of these degrees of freedom
up to substantial dopings. This may account for the presence of sizable spin excitations 
observed by inelastic neutron scattering up to large doping levels in ${\rm La_{2-x}Sr_xCuO_4}$ 
(LSCO) compounds \cite{wakimoto1,lipscombe,wakimoto2,yamada2}. 

The CO-QCP scenario was then extendend by relating the CO-QCP
with d-wave superconductivity and the pseudogap formation \cite{perali} and with their
isotopic dependencies \cite{andergassen}. The effects of critical CO fluctuations
on ARPES \cite{seibold}, optical \cite{enss} and Raman \cite{CDGS} spectra were subsequently explored.

We will show here that Raman spectra in LSCO
are consistent with the simultaneous presence of nearly-critical charge and spin fluctuations,
these latter displaying properties in agreement with the spectra observed by neutron experiments.

\section{The spin and charge collective modes}
The charge- and ``enslaved'' spin-fluctuation propagators near the CO-QCP 
has the generic form prediceted by the Hertz \cite{hertz} 
theory of quantum criticality with dynamical critical exponent $z=2$
\beq 
D_\alpha({\bf q}, \omega_n)=-\frac{g_\alpha^2}{m_\alpha+\nu_\alpha
({\bf q}-{\bf q}_{\alpha c})^2+|\omega_n|+
\omega_n^2/{\OO_\alpha}} \label{propagator}
\eeq
where $\alpha=c,s$ refers to charge or spin fluctuations.
The same form of the charge CM (but for the $\omega^2$ term in the denominator)
was microscopically calculated within a strongly-correlated 
Hubbard-Holstein model in Ref. \cite{CDG}, where the presence of a QCP around
optimal doping  was first predicted using realistic parameters for the cuprates.
For spin modes, Eq. (\ref{propagator}) has been customarily considered
in theories involving spin fluctuations \cite{chubukov}.
Here, $\omega_n$ are boson Matsubara frequencies, $g$ is the 
QP-CM coupling, $\nu$ is a fermion scale setting the rapid increase of the
denominator when the CM momenta move away from the critical ones. This 
dispersive term is assumed to have an intrinsic cutoff $\Lambda$ above which the
CM no longer exists. $m_c\propto \xi^{-2}$ is
proportional to the inverse square correlation length and it measures the distance 
to the CO transition. The mass $m_s$ of the ``enslaved'' spin CMs can follow the charge CM
mass, but no theory has yet been elaborated for this effect and in the following it will
be phenomenologically determined. 
In a recent work \cite{CGDE} we pointed out that 
the subleading $\omega^2$ term not only arises from higher-order contributions
of the electronic particle-hole excitations, but, in the case of charge modes, 
it relates fermions and phonon scales and encodes the dynamical nature of the phonons
thereby introducing the  cutoff $\OO_c$. 
Typical values were estimated to range from tens to several hundreds of cm$^{-1}$ \cite{CGDE}.
For spin fluctuations, a similar term arises from the coupling to higher-energy
propagating spin modes yielding a somewhat larger scale $\OO_s$ of
the order of thousends of cm$^{-1}$. 
While these terms play a relevant role in optical conductivity by setting the
momentum-dissipation scale \cite{CGDE}, here we point out a different physical
role by individually discussing the role of the different terms in the
denominator of Eq. (\ref{propagator}). The mass $m$ represents the minimal energy
required to create a fluctuation peaked around momenta $\qvec_\alpha$, 
while the linear frequency term extablishes
the diffusive (i.e. relaxational) character of these fluctuations arising from 
the damping due to decay into particle-hole pairs. The $\omega^2$ term, once it is
analitically continued to real frequencies, sets the propagating character
of the CM above the $\OO$ energy. The propagation then follows the
$\sqrt{\OO (m+\nu (\qvec-\qvec_\alpha)^2)}$ dispersion. Were it not
for the diffusive part of the CM spectrum, the propagating part would then occur
in the range between $\omega_1\approx \sqrt{\OO m}$ and $\omega_2\approx \sqrt{\OO 
\Lambda}$. In this case the spectral distribution of the modes closely
resembles the ``gapped marginal-FL'' one considered in Ref. \cite{normanchubukov}
to describe optical conductivity. The main advantage of the present form 
is that Eq. (\ref{propagator})
describes both a diffusive low-energy regime important for the leading critical
behavior and propagating behavior at intermediate-high energies relevant in
Raman and optical spectra. Moreover, the link between Eq. (\ref{propagator}) and
specific microscopic models allows for estimates of the relevant parameters
and for a transparent interpretations of
the physically relevant absorption mechanisms. 

\begin{figure}
\includegraphics[scale=0.3]{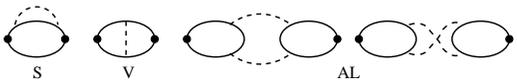}
\caption{Diagrams for the CM corrections to the optical or Raman response function. The full 
dots represent the current or Raman vertices. The solid  lines represent the QP propagator 
 and the dashed lines represent the interaction mediated 
by CO or AF CM. Self-energy, vertex, and Aslamazov-Larkin-like contributions
are labeled by S, V, and AL, respectively.}
\label{diagrams}
%
\end{figure}

In the following we will mostly focus on the Raman spectra to identify from symmetry
arguments the role of spin and charge fluctuations in the various Raman channels.
In this framework we will consider different processes reported in Fig. \ref{diagrams}
The S-V terms represent the perturbative ``dressing'' of the 
QPs by the CMs. On the other hand, the AL diagrams closely resemble those
of the Aslamazov-Larkin theory of paraconductivity due to gaussian Cooper-pair
fluctuations. 
Owing to their different physical character in Raman, we separate in the following 
discussion the two contributions.

\section{Anomalous Raman spectra: The charge gaussian contribution}
By suitably choosing the polarization of the incoming and outgoing photons in Raman scattering
experiments, momentum-dependent vertices are introduced in the response, which select 
the regions in momentum space
from where excitations are originated \cite{reframan}. As a consequence,
characteristic features arise  in the Raman response 
with distinct doping and polarization dependencies 
which in turn allow to extract valuable informations
about the excitations. This is the main issue
of this work, where we will consider the $B_{2g}$ channel 
with a vertex  $\gamma_{B_{2g}}=\sin k_x \sin k_y$ large along the diagonals of the
Brillouin zone, and the $B_{1g}$ channel  with a vertex $\gamma_{B_{1g}}=\cos k_x-\cos k_y$,
which is large around the ``hot'' region near the $(\pm \pi,0)$ and  $(0,\pm \pi)$ points.
In Ref. \cite{hacklPRL} it was found that Raman spectra display low-energy anomalous absorption
in the $B_{2g}$ channel 
at low doping $x<0.05$ and in the $B_{1g}$ channel at higher dopings.
These low-energy features were
attributed \cite{CDGS} to gaussian fluctuations of the charge density (quite similarly to the Cooper pair
gaussian fluctuations in the Aslamazov-Larkin theory of paraconductivity). Based on symmetry arguments 
it was also shown  that only charge CM could contribute with specific momentum structures,
which agree with the ones inferred from neutron-scattering experiments \cite{yamada}.
Quite remarkably, the anomalous spectral features were fitted at various temperatures by simply
adjusting the CM mass $m(x,T)$. This allowed us a straightforward determination of the 
T-dependence of $m$, which displayed the linear behavior  expected for a quantum-critical CM crossing
over below a temperature $T^*$ to a saturated nearly constant behavior typical of a state with
``quenched'' criticality with local-slow order (but without the long-range order, which would
correspond to $m=0$). This allowed to identify $T^*$ at two doping values, which (roughly) extrapolated
to zero [$T^*(x_c)=0$] at a critical doping $x_c$ to be identified with the QCP doping\cite{CDGS,hacklPRL}. 
These results were confirmed by a more
recent systematic analysis \cite{hackl}  at several other doping values. 
The recent analysis of the LSCO results includes samples at low and moderate 
doping levels having anomalous scattering in both the $B_{2g}$ and $B_{1g}$ channels, respectively. 
This allows us to draw some conclusions:
i) there is a crossover temperature $T^*$, clearly related to CO, which decreases upon
increasing doping. Its doping dependence closely follows the doping dependence of the crossover
temperature for the opening of the pseudogap. This allows us 
to clearly relate the latter one
to CO; ii) this conclusion is further strengthened by the observation that the the electronic DOS starts 
to be substantially reduced in the background spectra (i.e., in spectral ranges completely separated by the
anomalous features) below the same temperature $T^*$. This provides a completely independent determination
of $T^*$, which is consistent with the $T^*$ obtained from $m(x,T)$; 
iii) the extrapolated $T^*$ vanishes  around optimal doping at $0.16<x_c<0.19$, thereby
indicating this range to be the location of the CO-QCP.

We also notice that the analysis of gaussian CO fluctuations in the $B_{1g}$ Raman spectra allows us to
identify some key parameters of the CM in Eq. (\ref{propagator}). In particular the mass $m$ 
and the ``diffusivity'' scale $\OO$ can be estimated by considering the
low-energy side of the spectra. As we shall see in the next section, these quantities 
together with the high-energy cutoff $\Lambda$, determine the  spectral function of the CO-CM,
which can account for the overall shape of the $B_{2g}$ spectra.

\section{Disentangling spin and charge fluctuations from B$_{1g}$ and B$_{2g}$ spectra}
To analyze the whole spectra over a broader energy range, we need to consider physical processes
where the CMs dress the fermionic QPs. We accordingly define a 
non-resummed Raman response $\tilde \chi_{S-V}$ from the first two diagrams of Fig. 1. The presence of 
impurities can be described introducing the memory function $M_{S-V}=i\Gamma-\omega {\tilde \chi_{S-V}}/W$, 
where $W$ is the optical weight and $\Gamma$ is the impurity scattering rate. The full Raman response 
function is then found as
\beq
\chi_{S-V}=\frac{W\omega}{\omega+M_{S-V}(\omega)}.
\label{chi}
\eeq
The S-V Raman response function of the corresponding CM can be written as in Eq.(\ref{chi}).
 The memory function can be cast in the perturbative form \cite{normanchubukov}
\beqa
&Im&M_{S-V}(\omega)=\frac{g_\alpha^2}{\omega} \int_0^\infty dz \left[ \alpha^2 F(\omega)\right] 
\left[2\omega \coth\left(
\frac{z}{2T}\right) \right. \nonumber \\
 &-&\left.(z+\omega)\coth\left(\frac{z+\omega}{2T}\right) +(z-\omega)
\coth\left(\frac{z-\omega}{2T}\right) \right]
\eeqa
where T is the temperature and the CM spectral strength is
\beq
\alpha^2F(\omega)=\arctan\left(\frac{\Lambda\OO-\omega^2}{\OO\omega} \right)-
\arctan\left(\frac{m\OO-\omega^2}{\OO\omega} \right)
\eeq
 The real part of $M_{S-V}$ is found by Kramers-Kronig transformation. 

We are now in a position to make a simple working hypotesis: We assume that the CM's (both spin and charge)
provide the  interelectronic scattering needed to account for the whole Raman spectra over the entire
frequency range up to $\omega=8000$ cm$^{-1}$. The important step further is to recognize that the
momentum dependence of the CO and spin CM's is such that their {\it leading-order} contributions
via the diagrams S and V of Fig. 1, act separately in the different channels. Specifically one can
show that the (leading-order) contribution of the spin-CM only acts in the $B_{1g}$ channel, while it
is cancelled in the $B_{2g}$. The reverse is true for the CO-CM. This leading-order separation allows to
retrace back the difference in the spectra to differences of the CM producing the scattering in each channel.
Specifically, large values of $\OO$ results in diffusive-like modes as shown in the inset of Fig. 2(b),
while smaller values of $\OO$ give rise to more propagating modes with the square-like spectrum
typical of a marginal Fermi-liquid  displayed in the inset of Fig. 2(c).
We find it remarkable that the generic quantum-critical form of the CM propagators in Eq. 
(\ref{propagator}) can be tuned to display either the character of a diffusive critical mode
or that of a marginal-Fermi-liquid mediator.

\begin{figure}
\includegraphics[scale=1.4]{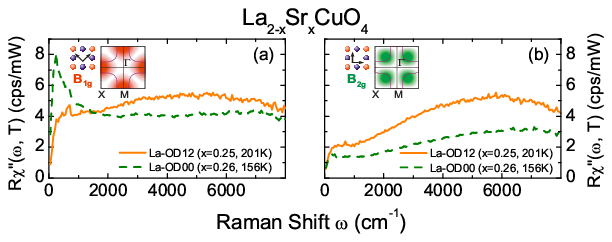}\vspace{0.5truecm}\\
\includegraphics[scale=0.23]{GRILLI-FIG2c.eps}\vspace{0.7truecm}\\
\includegraphics[scale=0.23]{GRILLI-FIG2d.eps}
\caption{Raman spectra in the $B_{1g}$ channel (a) and $B_{2g}$ channel (b)
for a LSCO sample at $x=0.25$ (solid curves) and $x=0.26$ (dashed curves).
(c) $B_{1g}$ Raman spectra calculated from spin CMs with 
the $\alpha^2F$ displayed in the inset
with $m_s=80 $, $\OO_s=10^5$,
$\Lambda_s=2000$, and $g_s^2=12.1$ (solid curve) and $m_s=500 $, $\OO_s=5\cdot 10^5$,
$\Lambda_s=2000$, and $g_s^2=7.3$ (dashed curve) (all energy units are in cm$^{-1}$). 
(d) $B_{2g}$ Raman spectra calculated from CO CMs with the $\alpha^2F$ displayed in 
the inset with $m_c=300$, $\OO_c=200$,
$\Lambda_c=15000$, and $g_c^2=5.5$ (solid curve) and $m_c=300$, $\OO_c=100$, and
$\Lambda_c=90000$  and $g_c^2=4.5$ (dashed curve) (all energy units are in cm$^{-1}$). 
}
\label{fig.4}
%
\end{figure}

Since we aim to keep the treatment as simple as possible and to carry out analytical calculations
as far as possible, we choose to insert free quasiparticles in the S-V diagrams thereby disregarding
the pseudogap effects on the fermions. This is obviously justified only at sufficiently large dopings.
This is why we consider the two ($B_{1g}$ and $B_{2g}$) spectra of two overdoped LSCO samples \cite{hackl}
reported in Figs. 2(a) and 2(b). 

The most obvious changes with doping are observed in the $B_{1g}$ spectra (Fig. 2(a)). 
While there is a well defined peak in the range 200 to 500 cm$^{-1}$ followed by a flat part up to 
8000 cm$^{-1}$ in the non-superconducting sample ($x=0.26$; dashed line), the low-energy peak 
is completely absent at $x=0.25$ ($T_c \approx 10K$; full line), and only a hump below 100 cm$^{-1}$ is left. 
The differences are more more subtle in $B_{2g}$ symmetry (Fig. 2 (b)). As a general feature 
there is a hump at approximately 500 cm$^{-1}$. At $x=0.25$ (full line) the subsequent increase occurs in 
a range of 3000 cm$^{-1}$ 1ending in a broad maximum at 5000 cm$^{-1}$. At $x=0.26$ (dashes) the increase is 
very smooth and there is no maximum any more in the observed range. 

We then notice that even at fixed doping the spectra look qualitatively
different in the two channels. In particular, for the  $x=0.25$ sample,
while the $B_{1g}$ spectrum has the rounded shape of a hump
plus a (slightly higher) hump, the $B_{2g}$ spectrum looks like a flattish step followed by a substantially
(nearly a factor two) higher hump. This 
behavior is also observed at lower doping levels \cite{hackl}. 
Starting from different shapes of the spin and charge CMs (see full curves in the insets) we reproduce quantitatively the spectra in the two channels. 
Of course the precise fitting of the spectra would require the adjustment of many parameters.
The resummed spectra derived from Eq. (\ref{chi}) depend on the optical weight $W$, on the 
elastic scattering from impurities $\Gamma$ as well as on the parameters of the CMs.
To extract qualitative informations on these latters we choose to keep $W$  and $\Gamma$
fixed trying to reproduce the semiquantitative behavior of the spectra by only
adjusting the CM parameters. Specifically we find that
the $B_{1g}$ spectrum arises from (spin) modes with a substantially diffusive form 
($\OO_s\approx 5\cdot 10^4$ cm$^{-1}$, $m_s\approx 80$ cm$^{-1}$) with a rapid rise of the spectrum
reaching a broad maximum around 500-700 cm$^{-1}$. The spin mode then ``dies'' above a frequency of 
about 2000 cm$^{-1}$. On the other hand, the $B_{2g}$ spectrum of the $x=0.25$ sample acquires the step+hump shape
if a (charge) CM is chosen with a markedly propagating
character [$\OO_c\approx 10^2$ cm$^{-1}$) and $m_s\approx 500$ cm$^{-1}$]. The low-energy weight of this
mode only arises from its small diffusivity and then substantially rises only around 300 cm$^{-1}$.
Also this mode no longer exists above a typical frequency of the order of 2000 cm$^{-1}$.

Upon slightly increasing the doping to $x=0.26$ the $B_{2g}$ (charge-related)
spectrum changes little in shape and is simply reduced in intensity. Also the related collective mode
stays propagating and it is just moderately reduced in intensity. To account for the data one only needs to
consider a broader spectral range extending up to 3000 cm$^{-1}$.
On the contrary, the spin-related  $B_{1g}$ spectrum changes remarkably in shape: It has a narrow
Drude-like form at low frequency and then flattens and stays nearly constant up to high frequencies.
This behavior is recovered in the theoretical spectrum by choosing an even more strongly
damped mode ($\OO_s\approx 10^5$ ) (which reduces the overall height of the spectrum)
and by further increasing the mass $m_s=500$ cm$^{-1}$. This increase lowers the spectral weight at low 
frequency thereby reducing the scattering due to this mode. In this way the narrow Drude-like peak is 
obtained. Then, as soon as the frequency increses, the broad spectrum of the mode enters into play
producing the flattening of the $B_{1g}$ spectrum.

\section{A brief analysis of optics}
\begin{figure}
\includegraphics[scale=0.2]{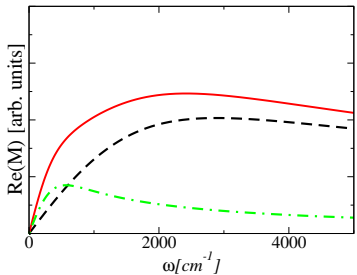}
\caption{Optical memory function (arbitrary units) obtained from the summed effect of a CO CM
with $m_c=30$, $\OO_c= 20$, and $\Lambda_c=2000$  (dot-dashed line) and a spin CM with 
$m_s=400$, $\OO_s= 2000$, and $\Lambda_s=2000$  (dashed line). 
The overall coupling factor are
$g^2_c=5$ and $g^2_s=1$ (all energy units in cm$^{-1}$)}
\label{optmem}
%
\end{figure}

The above analysis of Raman spectra exploited the Raman form factors to identify 
the role of spin and charge CMs in the various channels. This distinction is
no longer possible in optical conductivity, where all the different CMs
equally contribute to the spectra. Another remarkable difference is that 
optical spectra strongly depend on the momentum-charge conservation laws, which 
may introduce additional energy scales related to the dissipation mechanisms
\cite{CGDE}. As a consequence of the conservation laws, all the contributions 
from the diagrams of Fig. \ref{diagrams} are tightly related and implement 
conspicuous concellations in the case of nearly clean systems. Here we will 
instead neglect these effects assuming that the momentum conservation is 
substantially violated both from static, Umklapp, and dynamical dissipation mechanisms.
In this case the S-V and AL contributions in the current-current
response no longer display severe cancellations and one can obtain a semiquantitative
estimate of optical spectra from the S-V contributions only. This is what we
carry out in this section by replacing the Raman response $\tilde \chi_{S-V}$ entering
the memeory function with the current-current response  $\tilde \chi_{jj}$ thereby obtaining
the optical memory function $M_{opt}=i\Gamma-\omega {\tilde \chi_{jj}}/W$.
We then are in the position to qualitatively compare the contribution of CO {\it and} spin CM (they act 
on equal footing in optics)
to the optical memory function recently obtained in experiments \cite{vandermarel}.
For an illustrative purpose we calculate the optical memory function by simply adding 
the effects of the CO and spin CMs. The result is reported in Fig. \ref{optmem},
where the memory function of a spin CM (black curve) with diffusive
character and moderate mass is displayed together with
the memory function of a CO-CM with smaller mass and a more propagating character.
Clearly the resulting total memory function (red curve) is similar to the
ome reported in Fig. 1 of Ref. \cite{vandermarel} for an optimally doped 
HgBa$_2$CuO$_{4=\delta}$ sample in the normal state
well above $T_c$. The purpose here is not to carry out a detailed 
quantitative comparison, but simply to show that our scheme based on the
coexistance of spin and CO modes may account for optical data as well. We also
notice in passing that our scenario in principle is not in contradiction with the one
presented in Ref. \cite{vandermarel}, where two markedly distinct features were
deduced in the bosonic spectra. In our case the low-energy bosonic part of the
spectrum would correspond to the more propagating CO modes, while the broader
part of the bosonic spectrum at higher energy should be identified with the
spin diffusive modes.

\section{Discussion and conclusions}
The results reported in the previous sections are relevant for several reasons.
First of all, the very fact that the spectra in the various channels 
are different and can be accounted for by scattering due to CMs with different
momentum structures, is an indication that the physics of the cuprates
is not ruled by zero momentum modes or by local nearly structureless modes.
Then one is led to conclude that i) retarded interactions with momentum structure  
at finite wavevectors are at work. ii) The temperature dependence of (the mass of) 
these modes is 
relevant to account for the temperature dependence of the low-energy anomalous 
features. Moreover if one keeps the mass of the modes (i.e. their
typical low energy scale) temperature independent, the resulting spectra
depend too strongly from temperature (even at large frequencies) in comparison to the
experimental Raman spectra. This arises because the bosonic
caracter of the modes introduces Bose distribution functions, which substantially
depend on temperature \cite{normanchubukov,mosca}. iii) To recover the (near) temperature independence 
of the observed spectra at high frequency  one has to keep
the precise temperature dependence of the mass \cite{mosca}. 
We also found that both charge and spin modes
contribute to scattering allover the frequency range. Our leading-order 
symmetry arguments then allowed us to identify the contribution of each mode
in the different channels. However, one should also keep in mind that 
the pretemption to account for the shape of the spectra keeping separate the spin  and 
charge contribution looses its validity at energies as high as several thousends
of cm$^{-1}$. The symmetry arguments based on the critical structure (i.e. strongly momentum
dependent at low energies) are no longer justified at such high energy and only 
provide rough indications.
Moreover, not only the symmetry arguments to distinguish spin and charge contribution
become unreliable, but also the physical distinction between these modes
is questionable: Indeed 
a strong mixing of the spin and charge degrees of freedon is observed in LSCO above
this energy scale \cite{doubleFS}.
In the light of all these consideration one can then understand why 
above 4000 cm$^{-1}$ the spectra in both channels become quite
similar signaling that the distinction between spin and CO CM breaks down at high energy.

Nevertheless we find remarkable that the differences in the spectra are understood in terms
of scattering due to CMs with different characteristics. Namely, the spin modes are strongly
overdamped and rapidly loose intensity upon increasing doping. This behavior is also
clearly observed in inelastic neutron scattering \cite{wakimoto1,lipscombe,wakimoto2}.
In particular, we find that the strong reduction (by nearly a factor of two in the spin DOS)
in going from $x=0.25$ to $x=0.26$ is in qualitative agreement with a similar reduction
observed in the neutron spectra of Ref. \cite{wakimoto2} in passing from $x=0.25$ to $x=0.27$.
We also find it quite reasonable that in the same doping range, the CO-CMs are more stable upon
doping and are more propagating (i.e. less damped). This is what is expected for  critical
modes which are much closer to their QCP (which we locate around $x=0.18$ from the low-energy
anomalies).


\end{document}